\documentclass[a4paper,11pt,DIV12]{scrartcl}

\addtokomafont{disposition}{\rmfamily}

\usepackage[utf8]{inputenc}
\usepackage{enumitem}
\usepackage{amsmath}
\usepackage{amssymb}
\usepackage{latexsym}
\usepackage{microtype}
\usepackage{xspace}
\usepackage{hyperref}
\usepackage{xcolor}
\usepackage{tikz}
\usetikzlibrary{patterns,snakes}
\usepackage{ifthen}
\usepackage{orcidlink}

\newtheorem{theorem}{Theorem}

\newtheorem{proposition}{Proposition}
\newtheorem{lemma}{Lemma}

\newtheorem{remark}{Remark}
\newtheorem{example}{Example}

\newenvironment{proof}[1][]{\pagebreak[3]\noindent\textit{Proof\ifthenelse{\equal{#1}{}}{}{ (#1)}: }}{\pagebreak[3]\medskip}

\setlist{noitemsep}

\newcommand{\N}{\mathbb{N}}

\newcommand{\Oh}{\mathcal{O}}

\newcommand{\cP}{\mathcal{P}}
\newcommand{\cQ}{\mathcal{Q}}

\newcommand{\set}[2]{\left\{\hspace*{0.5pt}#1\vphantom{#2}\hspace*{0.5pt} \left|\hspace*{2pt} \vphantom{#1}#2 \right.\hspace*{0.5pt}\right\}}
\newcommand{\os}[1]{\left\{\hspace*{0.5pt}#1\hspace*{0.5pt}\right\}}
\newcommand{\smallset}[1]{\left\{#1\right\}}

\newcommand{\abs}[1]{\left|\mathinner{#1}\right|}

\newcommand{\ceil}[1]{\left\lceil\mathinner{#1} \right\rceil}

\newcommand{\Even}{\textsc{Even}\xspace}
\newcommand{\Odd}{\textsc{Odd}\xspace}
\newcommand{\attr}{\mathop{\mathrm{attr}}}
\newcommand{\reset}{\mathop{\mathrm{reset}}}

\newcommand{\eex}{\hspace*{\fill}$\Diamond$}
\newcommand{\qed}{\hspace*{\fill}$\Box$}

\newlength{\msize}
\newlength{\isep}
\newlength{\osep}
\tikzset{positionStyle/.append style={draw,minimum size=\msize,inner sep=\isep,outer sep=\osep}}

\title{Reachability Games and Parity Games} 

\author{Volker Diekert\hspace*{2pt}\orcidlink{0000-0002-5994-3762} \and Manfred Kuf\-leitner\hspace*{2pt}\orcidlink{0000-0003-3869-416X}}
  
\date{University of Stuttgart, FMI, Germany\\
\texttt{\{diekert,kufleitner\}\@fmi.uni-stuttgart.de}}

\begin{document}
\maketitle

\begin{abstract}
\noindent
\textbf{Abstract.}\ 
Parity games are positionally determined. This is a fundamental and classical result. In 2010, Calude et al.{} showed a breakthrough result for finite parity games: the winning regions and their positional winning strategies can be computed in quasi-polynomial time. 

In the present paper we give a self-contained and detailed proofs for both results. 
The results in this paper are not meant to be original. The positional determinacy result is shown for possibly infinite parity games using the ideas of Zielonka which he published in 1998. In order to show quasi-polynomial time, we follow Lehtinen's register games, which she introduced in 2018. Although the time complexity of 
Lehtinen's algorithm is not optimal, register games are conceptually simple and 
interesting in their own right. Various of our proofs are either new or simplifications of the original proofs. The topics in this paper include the definition and the computation of optimal attractors for reachability games, too. 
\end{abstract}

\tableofcontents

\section{Introduction}

A game on a graph is played by two players who move from one vertex to another. The vertices are often called \emph{positions}. Every move needs to follow an edge in the graph. Each position belongs to one of the players and the owner of the position chooses the next move. The resulting sequence of moves can be finite or infinite. Basically, there can be two reasons for a game to be finite: the game ends in a sink (i.e., a vertex where no moves are possible) or one of the players has won the game. The other situation is that the game continues indefinitely. Infinite games also have a winner; the winner depends on the sequence of vertices visited during the game or, alternatively, on the sequence of moves (i.e., edges) taken by the players. When considering infinite duration games, then a typical approach in the literature is to avoid finite games by disallowing both sinks and finite winning sequences. In this paper, we take a slightly different approach. We allow game graphs to be infinite and to have sinks, and we consider winning conditions which allow both finite and infinite games. This way, we are able to discuss reachability games and parity games in a uniform way. 

A \emph{strategy} is a rule for choosing a player's next move; the chosen move can depend on the current position (i.e., the current vertex in the graph) and all previous moves. Players \emph{follow} a strategy if, whenever it is their turn, they always use the strategy's suggestion as their next move. A strategy is \emph{winning} if, by following the strategy, the player wins against all possible replies of the opponent. This depends on the starting position; there might be some starting positions where the strategy is winning and others where it is not winning. A game is \emph{determined} if for every starting position exactly one of the players has a winning strategy. Not all games are determined; the example by Gale and Stewart of a non-determined game relies on the axiom of choice~\cite{GaleStewart1953}.

After introducing a general framework for games on graphs, we consider reachability games and parity games in more depth. The objective for one of the players in a \emph{reachability game} is to eventually visit a position in a given target set~$R$; the objective of the other player is to never visit a position in $R$. In a \emph{parity game}, there is a finite set of non-negative integers and each vertex is colored with one of these integers; the colors are also called \emph{priorities}. A game which ends in a sink is losing for the owner of the sink (i.e., a player loses immediately if they cannot move); all other games are infinite. In an infinite game, the largest color which is seen infinitely often determines the winner. One player wins if this color is even and the other player wins if it is odd. Among the numerous applications of parity games, we mention the following two: parity games play an important role in model-checking modal $\mu$-calculus \cite[Part V]{LNCS2500automata}, and they can be used for proving the complementation lemma in Rabin's Tree Theorem~\cite{rab69}; see e.g.~\cite{Tho97handbook}.

Martin's 
Determinacy Theorem shows that if the winning condition in a game on graphs is a Borel set, then the game is determined~\cite{Martin1975}. This includes both reachability games and parity games. However, the winning strategies from the Borel Determinacy Theorem need to store all the previous moves. Gurevich and Harrington proved that \emph{finite-memory strategies} suffice for parity games over finite game graphs~\cite{GurevichH82stoc}: at every starting position, exactly one of the players has a winning strategy which only takes into account the current position and a fixed number of bits of information about the past (and this fixed number of bits can be updated move by move). Independently of one another, Emerson and Jutla~\cite{EmersonJ1991focs} and Mostowski~\cite{Mostowski1991tr} further improved this result by showing \emph{positional determinacy} (or \emph{memoryless determinacy}) of finite parity games. In a \emph{positional} strategy, the next move only depends on the current position. Positional strategies are also known as \emph{memoryless}. Positional determinacy means that, for every starting position, exactly one of the players has a positional winning strategy. Zielonka showed memoryless determinacy for infinite parity games in which every vertex has only a finite number of successors~\cite{zielonka98tcs}, but he also observed that only some minor adjustments are necessary to generalize this result to arbitrary infinite graphs. Therefore we consider parity games over arbitrary graphs confirming his observation. The present proof is based on notes of the first author when he attended a lecture by Zielonka held in Paris on January~19th,~1996. As a tool for our proof, we show that reachability games are positionally determined. The result is well-known and considered to be folklore. For the sake of completeness, we include the proof.

Algorithmically solving a game usually means one of two things. Firstly, given a starting position, one wants to know the winner of the game (i.e., the player with a winning strategy). And secondly, we can solve a game by computing winning regions and winning strategies for the two players. Since a solution to the first problem typically also involves the computation of a winning strategy, the two problems are equivalent in practice. We only consider solutions of games with finite game graphs. It is folklore that reachability games can be solved in time $\Oh(n+m)$ for game graphs with $n$ vertices and $m$ edges; see e.g.~\cite[Exercise~2.6]{LNCS2500automata}. We give a version of this algorithm which computes \emph{optimal} strategies. There is a large and increasing number of algorithms for solving parity games; we refer to the Oink project by van Dijk~\cite{vanDijk2018tacas} for an overview. Nowadays, Zielonka's algorithm~\cite{zielonka98tcs} is considered to be the most classical one. It is relatively easy to describe and it is often fast in practice~\cite{FriedmannL2009atva}. However, Friedmann showed that there are instances where Zielonka's algorithm uses an exponential number of steps~\cite{Friedmann2011rairo}. In the same paper, Friedmann gave an upper bound of $\Oh(n^d)$ on the number of recursive calls in Zielonka's algorithms for parity games with $n$ vertices and $d$ colors. Since every recursive call involves solving two reachability games, this yields a running time of $\Oh(n^d(n+m)) \subseteq \Oh(n^{d+2})$. We give an analysis of Zielonka's algorithm which shows that its running time is in $\Oh(n^{d-1}(n+m)) \subseteq \Oh(n^{d+1})$. 

Calude et al.~\cite{CaludeJKLWS2017stoc} showed that parity games with $n$ vertices can be solved in quasi-polynomial time $2^{\log^{\Oh(1)}(n)}$. This led to a series of quasi-polynomial algorithms; see e.g.~\cite{LehtinenPSW2022lmcs,Parys2019} for brief overviews. We give a version of Lehtinen's algorithm~\cite{Lehtinen18}. Her quasi-polynomial time algorithm is conceptually simpler than the algorithm by Calude et al.{} but less efficient in the worst case. Lehtinen's algorithm uses Zielonka's algorithm 
on a larger game graph but with $2+2\ceil{\log_2 n}$ colors, only. 
The resulting running time is $n^{\Oh(\log n)} d^{\Oh(\log^2 n)}$ for a game with $n$ vertices and largest color $d$. This is 
not optimal. 
For instance, the recent modification of Zielonka's algorithm~\cite[Theorem~3.3]{LehtinenPSW2022lmcs} has a running time of $\Oh\left(n^{6.9+2 \log \left( 1+\frac{d}{2 \log n} \right)} \right)$.

As usual, we use \emph{random access machines} to measure the time complexity of algorithms; see e.g.~\cite[Chapter~2.2]{CLRS22}.

\section{Games on Graphs}

A \emph{game graph} $G = (V_0, V_1, E)$ is a directed graph such that the \emph{vertices} $V = V_0 \cup V_1$ are partitioned into two sets $V_0$ and $V_1$ with $V_0 \cap V_1 = \emptyset$. We allow $V$ to be infinite. The set of \emph{edges} is $E \subseteq V \times V$. Depending on the setting, the game graph might have additional information such as labeled edges or a coloring of the vertices. A \emph{sink} is a vertex $v \in V$ without outgoing edges. The set of all finite paths in the graph $(V,E)$ is denoted by $E^*$; and $E^\infty$ is the set of all finite of infinite paths. We consider $E^\infty$ to be a subset of $V^+ \cup V^\omega$, i.e., a path $\alpha \in E^\infty$ is either a non-empty finite sequence $\alpha = v_1 \cdots v_k$ or an infinite sequence $\alpha = v_1 v_2 \cdots$ of vertices $v_j$ such that any two consecutive vertices $v_j, v_{j+1}$ satisfying the edge relation $(v_j, v_{j+1}) \in E$. Similarly, we have $E^* \subseteq V^+$. There are two \emph{players}, player $0$ and player $1$. The vertices in $V_i$ \emph{belong} to player $i \in \os{0,1}$. A \emph{position} is a vertex $u \in V$. At position $u$, player $i \in \os{0,1}$ with $u \in V_{i}$ chooses $v \in V$ with $(u,v) \in E$. The next position is $v$ and the game continues at this position. This is called a \emph{move} of player $i$.
We use the term \emph{position} rather than vertex for an element $v \in V$ to emphasize that $v$ is part of a sequence of moves; on the other hand, for graph properties such as paths we use the term \emph{vertex}.

A set of \emph{games} $C$ is a subset of $E^\infty$ such that every path in $E^\infty$ has a unique prefix in $C$. This prefix does not need to be proper. Note that no path in $C$ has a proper prefix in $C$; i.e., either a game is infinite or the game immediately ends as soon as a finite sequence of moves defines a path in $C$.
A \emph{winning condition} is a partition $C = C_0 \cup C_1$. Here, $C_i$ is the set of games which player $i$ wins, and we have $C_0 \cap C_1 = \emptyset$.
The winning condition $C_i$ of player $i$ does \emph{not depend on finite prefixes} if for every $p\beta \in C$, we have that $\beta \in C_i$ implies $p\beta \in C_i$. We note that this property of $C_i$ also depends on $C_{1-i}$ because we consider all games in $C = C_0 \cup C_1$.
A \emph{game} $(G,u,C_0,C_1)$ consists of a game graph $G$, an initial position $u \in V$, and a winning condition $C_0, C_1$. The latter gives the set of games as $C = C_0 \cup C_1$. If the winning condition is clear from the context, then the game is denoted by $(G,u)$; if $G = (V_0, V_1, E)$, then we usually do not distinguish between $(G,u)$ and $(V_0, V_1, E, u)$ in the sense that both describe the same game. Remember that, by abuse of notation, a \emph{game} is also a sequence in $C$.
A game (in the sense of a sequence in $C$) on $(G,u)$ is created by moves of the players, starting at position $u$; the game is either infinitely long, or it finishes immediately if the current sequence of positions yields a path in $C \cap E^*$. Note that every game has a unique winner $i \in \os{0,1}$ because if the game reaches a sink (in particular, there was no previous point at which a player has won the game), then the path leading to this sink needs to be in $C$.

Intuitively, a strategy for player $i$ defines its next move; this move can depend on the current position $v_k \in V_i$ as well as the sequence of the previous positions $v_1,\ldots,v_{k-1}$. More formally, it is a partial map $\sigma : E^* \to V$ such that after the moves $v_1 \cdots v_k$ with $v_k \in V_i$, the next move of player $i$ is $v_{k+1} = \sigma(v_1 \cdots v_k)$. In particular, the strategy is required to satisfy $(v_k,v_{k+1}) \in E$. It does not have to be defined on all paths in $V^*V_i$ because some configurations might not be reachable if player $i$ always moves according to the strategy. Moreover, we sometimes do not care for certain positions how player $i$ moves. This is the case if all moves at this position lead to winning games, or if all games at this position immediately end.
In this paper, we are mostly interested in strategies which do not take the previous moves into account and only depend on the current position: A \emph{positional strategy} for player $i$ is a partial map $\sigma : W \to W$ for $W \subseteq V$ such that if $\sigma(w)$ is defined, then $w \in W \cap V_i$ and $(w,\sigma(w)) \in E$; i.e., the strategy only suggests legal moves and only for player $i$.
The set $W$ is called the \emph{support} of~$\sigma$.
A path $\alpha = v_1 v_2 \cdots \in E^\infty$ with $v_1 \in W$ follows a strategy $\sigma : W \to W$ for player $i$ if for all prefixes $v_1 \cdots v_j v_{j+1}$ of $\alpha$ such that $\sigma(v_j)$ is defined, we have $v_{j+1} = \sigma(v_j)$. In other words, whenever possible, player $i$ applies $\sigma$ for choosing their next move. Remember that if $\sigma(v_j)$ is defined, then $v_j \in W \cap V_i$.
Note that we allow $\alpha$ to leave and re-enter $W$.
A positional strategy $\sigma : W \to W$ for player~$i$ is \emph{winning} if all games $\alpha$ which start in $W$ and follow $\sigma$ are in $C_i$. In this case, we say that $\sigma$ is a \emph{$i$-strategy}.
If $\sigma : W \to W$ is an $i$-strategy and $\sigma(w)$ is not defined for $w \in W \cap V_i$, then all games starting in $W$, following $i$, and visiting $w$ at some point are winning for player $i$. This means that the only reason for $\sigma$ to be undefined at a position $w \in W \cap V_i$ is that the choice of the next move does not matter. For instance, this is the case if player $i$ wins as soon as the position $w$ is reached.
If $\sigma_0 : W_0 \to W_0$ is a $0$-strategy and $\sigma_1 : W_1 \to W_1$ is a $1$-strategy, then $W_0 \cap W_1 = \emptyset$; otherwise, a game starting in $W_0 \cap W_1$ and following both strategies would be winning for both players.

We can identify a positional strategy $\sigma : W \to W$ with the subgraph $(W,F)$ where the set of edges is $F = \set{(w,\sigma(w))}{\sigma(w) \text{ is defined}}$; note that $F \subseteq E$. 
Similarly, a positional strategy can be built-in into the game graph by replacing the edges $E$ by $E' = F \cup \set{(u,v) \in E}{\sigma(u) \text{ is not defined}}$.
We order $i$-strategies by $(W',F') \leq (W,F)$ if $W' \subseteq W$ and $F' \subseteq F$.
By Zorn's Lemma, there exist maximal $i$-strategies.
Maximal $i$-strategies $(W,F)$ have the following property: Whenever, for a position $w \in W \cap V_i$, the set of neighbors $\set{v \in W}{(w,v) \in E}$ in $W$ is nonempty, then there exists an edge $(w,v) \in F$. Otherwise, all possible $(W,F)$-following continuations (even those leaving $W$) from $w$ would lead to winning games for player $i$. By adding an edge $(w,v)$ to $F$ which stays inside $W$, the resulting strategy  would generate a subset of those continuations but with more edges, thereby contradicting the maximality of $(W,F)$. Therefore, maximal strategies always choose a move except if $w \in W \cap V_i$ is a sink or if all outgoing edges leave $W$.

\begin{proposition}\label{prp:unisup}
If the winning condition $C_i$ does not depend on finite prefixes, then the support of maximal $i$-strategies is unique. This means, if $(W_1, F_1)$ is maximal and $(W_2, F_2)$ is an arbitrary $i$-strategy, then we have $W_2 \subseteq W_1$.
\end{proposition}

\begin{proof}
Let $(W_1, F_1)$ be maximal and $(W_2, F_2)$ be an arbitrary $i$-strategy.
Let $W_3 = W_1 \cup W_2$ and $F_3 = F_1 \cup \set{(u, v) \in F_2}{u \in W_2 \setminus W_1}$, i.e., at positions in $W_1 \cap W_2$, we give preference to the strategy $(W_1,F_1)$. Consider a game $\alpha$ which starts in $W_3$ and follows $(W_3,F_3)$. If $\alpha$ never visits a position in $W_1$, then $\alpha$ follows the $i$-strategy $(W_2,F_2)$. Hence, $\alpha \in C_i$ in this case. If $\alpha = p \beta$ such that $\beta$ starts at a position in $W_1$, then $\beta$ follows the $i$-strategy $(W_1,F_1)$. As before, we see that $\beta \in C_i$. Since the winning condition does not depend on finite prefixes, we have $\alpha = p \beta \in C_i$. Thus, $(W_3,F_3)$ is an $i$-strategy. By maximality of $(W_1,F_1)$ and since $(W_1,F_1) \leq (W_3,F_3)$, we have $(W_1,F_1) = (W_3,F_3)$ and thus $W_2 \subseteq W_3 = W_1$.
\qed
\end{proof}

In the above situation, the support of a maximal $i$-strategy is called the \emph{winning region} of player $i$.
A game is \emph{positionally determined} if, for $i \in \os{0,1}$, there exist $i$-strategies $(W_i,F_i)$ such that $V = W_0 \cup W_1$. Since $W_0$ and $W_1$ are the supports of winning strategies for different players, we have $W_0 \cap W_1 = \emptyset$. Positional determinacy is also known as \emph{memoryless determinacy} in the literature.
If a game $G$ is positionally determined and player $i$ has an arbitrary (not necessarily positional) winning strategy for $(G,v)$, then $i$ also has a positional winning strategy for $(G,v)$: the opponent $1-i$ cannot win $(G,v)$ with a positional strategy against player $i$'s arbitrary strategy. Since $G$ is positionally determined, one of the players has a positional winning strategy for $(G,v)$; and this has to be player $i$ because it is not player $1-i$. By \emph{solving} a game $(G,v)$, we mean deciding which player has a winning strategy (if it exists at all); if $G$ is positionally determined, then exactly one of the players has a positional winning strategy (and the other player loses no matter which strategy they use).

\begin{remark}
  Sometimes one needs to distinguish whether there is one winning strategy for all starting positions in the winning region or whether for every starting position in the winning region there exists an individual winning strategy. Proposition~\ref{prp:unisup} shows that these two properties are equivalent for positional strategies and winning conditions which do not depend on finite prefixes.
  \eex
\end{remark}

\section{Reachability Games and Attractors}

Let $G = (V_0, V_1, E)$ be a game graph, and let $V = V_0 \cup V_1$.
Let $M \subseteq E^\infty$ be the paths which cannot be extended to the right (the identifier $M$ is for \emph{maximal} paths). In other words, we have $\alpha \in M$ if either $\alpha$ is infinite or $\alpha$ ends in a sink. In a \emph{reachability game}, the objective of one of the players is to reach a position in $R \subseteq V$. Suppose that player $i$ wins at all positions in $R$ in which case we call $R$ the \emph{target set} of player $i$. More formally, the winning condition for player $i$ is $C_i = (V \setminus R)^* R \cap E^*$, and the winning condition for player $1-i$ is $C_{1-i} = M \cap (V \setminus R)^\infty$. The winning conditions does not depend on finite prefixes.

\begin{theorem}\label{thm:reachdet}
  Reachability games are positionally determined.
\end{theorem}

\begin{proof}
  Let $(W_i,F_i)$ be a maximal $i$-strategy. Note that $R \subseteq W_i$. Let $W_{1-i} = V \setminus W_i$. It remains to show that player $1-i$ has a positional winning strategy with support $W_{1-i}$. If $u \in W_{1-i}\cap V_i$, then there exists no edge $(u,v) \in E$ with $v \in W_i$; otherwise $(W_i \cup \smallset{u}, F_i \cup \smallset{(u,v})$ is a bigger $i$-strategy than $(W_i,F_i)$. Next, we consider $u \in W_{1-i} \cap V_{1-i}$. If $u$ is a sink, then all games ending in $u$ are winning for player $1-i$ since it is impossible to reach a position in $R$. If $u$ is not a sink, then there exists an edge $(u,v) \in E$ with $v \in W_{1-i}$; otherwise $(W_i \cup \smallset{v}, F_i)$ is a bigger $i$-strategy than $(W_i,F_i)$. We set $\sigma(u) = v$. This leads to a strategy $\sigma : W_{1-i} \to W_{1-i}$ such that $\sigma(u)$ is defined for all $u \in W_{1-i} \cap V_{1-i}$ which are not a sink. Every game $\alpha$ starting in $W_{1-i}$ and following $\sigma$ never leaves $W_{1-i}$. It follows that $\alpha$ cannot enter a position in $R$ and, hence, $\alpha \in C_{1-i}$.
  \qed
\end{proof}

\begin{example}
  We consider the following reachability game with vertices $V_0 = \os{a,c,e}$ and $V_1=\os{b,d,f}$, edges $E=\os{ad,da,be,eb,de,ed,bc,ef,fc}$ where $xy$ denotes the pair $(x,y)$, and target set $R=\os{c,f}$ of player $0$. A graphical representation is
  
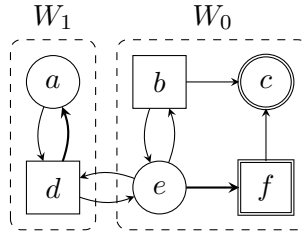
\begin{center}
  \begin{tikzpicture}[scale=1.4]
    \node[circle,positionStyle] (a) at (0,1) {$a$};
    \node[rectangle,positionStyle] (b) at (1,1) {$b$};
    \node[double,circle,positionStyle] (c) at (2,1) {$c$};
    \node[rectangle,positionStyle] (d) at (0,0) {$d$};
    \node[circle,positionStyle] (e) at (1,0) {$e$};
    \node[double,rectangle,positionStyle] (f) at (2,0) {$f$};
    \draw[-stealth,bend right=20] (a) edge (d);
    \draw[-stealth,bend right=20,thick] (d) edge (a);
    \draw[-stealth,bend right=20] (b) edge (e);
    \draw[-stealth,bend right=20] (e) edge (b);
    \draw[-stealth,bend right=20] (d) edge (e);
    \draw[-stealth,bend right=20] (e) edge (d);
    \draw[-stealth] (b) edge (c);
    \draw[-stealth,thick] (e) edge (f);
    \draw[-stealth] (f) edge (c);
    \draw[dashed,rounded corners] (-0.4,-0.4) rectangle (0.4,1.4);
    \draw[dashed,rounded corners] (0.6,-0.4) rectangle (2.4,1.4);
    \draw (0,1.4) node[above] {$W_1$};
    \draw (1.5,1.4) node[above] {$W_0$};
  \end{tikzpicture}
  \end{center}
  Round vertices belong to player $0$ and square vertices belong to player $1$. Double borders are used for states in the target set. The winning regions are $W_0 = \os{b,c,e,f}$ and $W_1 = \os{a,d}$ and the (in this case unique) positional strategies are $(W_i,F_i)$ with $F_0 = \smallset{da}$ and $F_1=\smallset{ef}$, indicated by thicker arrows.
  \eex
\end{example}

The proof of Theorem~\ref{thm:reachdet} suggests that the winning regions of the two players can be defined more explicitly.
A set of vertices $A \subseteq V$ is \emph{$i$-attracting} if the following two conditions hold:
\begin{enumerate}
\item If $u \in V_i$ and there exists an edge $(u,v) \in E$ with $v \in A$, then $u \in A$.
\item If $u \in V_{1-i}$ is not a sink and all edges $(u,v) \in E$ satisfy $v \in A$, then $u \in A$.
\end{enumerate}
The \emph{$i$-attractor} of $R \subseteq V$ is the smallest set of vertices which is $i$-attracting and contains~$R$. It is well-defined because $V$ is $i$-attracting, and the intersection of all sets which both contain~$R$ and are $i$-attracting also satisfies both properties. The $i$-attractor of~$R$ in a game graph $G$ is denoted by $\attr_i(G,R)$.

\begin{proposition}\label{prp:attr}
  Let $(W,F)$ be a maximal $i$-strategy for reaching the target set~$R$ within a game graph $G$. Then we have $W = \attr_i(G,R)$.
\end{proposition}

\begin{proof}
  Let $A = \attr_i(G,R)$ and $B = V \setminus A$.
  Since $(W,F)$ is maximal, the set~$W$ is $i$-attracting and it contains $R$; see the proof of Theorem~\ref{thm:reachdet} for details. 
  This shows $A \subseteq W$. For the converse, we show~$B \subseteq V \setminus W$ by giving a $(1-i)$-strategy $\sigma : B \to B$. Consider a position $u \in B$. We have $u \not\in R$ because $R \subseteq A$. In particular, if $u$ is a sink, then reaching $u$ is winning for player $1-i$. Therefore, we can assume that $u$ is not a sink. If $u \in V_i$, then there is no edge $(u,v) \in E$ with $v \in A$ because $A$ is $i$-attracting. Similarly, if $u \in V_{1-i}$, then there exists an edge $(u,v) \in E$ with $v \not\in A$ and we can set $\sigma(u) = v$. Every game which starts in $B$ and follows the strategy $\sigma$ stays in~$B$. Therefore, player $1-i$ wins at all positions in $B$ since he can avoid reaching a position in $R$. This shows $B \subseteq V \setminus W$ and, hence, $W \subseteq A$.
  \qed
\end{proof}

A consequence of Theorem~\ref{thm:reachdet} and Proposition~\ref{prp:attr} is that player $1-i$ wins at all positions in $V \setminus \attr_i(G,R)$.

\section{Optimal Strategies for Reachability Games}

In this section, we consider reachability games where the set of positions $V$ is finite. We still assume that player $i$'s winning objective is to reach a position in $R \subseteq V$. Let $(W_i, F_i)$ be a maximal $i$-strategy. Then for all starting positions $u \in W_i$, there is a maximal number of moves which are necessary for games which follows $(W_i,F_i)$ to reach a position in $R$. We are interested in a strategy for player~$i$ which minimizes this number. We do not aim at optimizing maximal $(1-i)$-strategies $(W_{1-i},F_{1-i})$ because usually games starting in $W_{1-i}$ and following $(W_{1-i},F_{1-i})$ are infinite; the only other case is that the game ends in a sink and if we would want to optimize the number of moves when targeting a sink, we could apply the same algorithms as below for player $1-i$ reaching this set of sinks.

The \emph{winning distance} is a function $d : V \to \N \cup \smallset{\infty}$ defined by:
\begin{itemize}
\item $d(u) = 0$ \;for $u \in R$,
\item $d(u) = \min\set{1+d(v)}{(u,v) \in E}$ \;for $u \in V_i \setminus R$,
\item $d(u) = \max\set{1+d(v)}{(u,v) \in E}$ \;for $u \in V_{1-i} \setminus R$.
\end{itemize}
Here, we let $\min \emptyset = \max \emptyset = \infty$ and $1 + \infty = \infty$.
A similar concept as the winning distance is the \emph{rank} of a position which is only defined for elements of the attractor~\cite[p.146]{zielonka98tcs}.
Since $d$ occurs on both sides of the definition, we need to show that the winning distance is well-defined.

\begin{lemma}\label{lem:windist}
  The winning distance is unique and well-defined. Moreover, we have $d(u) < \infty$ if and only if $u \in W_i$.
\end{lemma}

\begin{proof}
  A straightforward induction on $n \in \N$ shows that if $d(u) \leq n$, then $u \in W_i$. Next, we show that $u \in W_i$ implies $d(u) < \infty$.
  Since $V$ is finite, the $i$-attractor of $R$ can be computed by starting with $R$ and successively adding positions which contradict the current set to be $i$-attracting; this is repeated until the set does not change anymore.
  By using this naive algorithm, every time we add a position~$u$, we can define $d(u) < \infty$ (and possibly update previously defined values $d(v)$; these updates only affects positions $v$ satisfying $d(v) > d(u)$ before and after the update). Since $W_i = \attr_i(G,R)$, we have $d(u) < \infty$ for all $u \in W_i$. This concludes the second part of the lemma. Moreover, we have shown that there exists at least one winning distance.
  
  Suppose that $d$ and $d'$ are two different winning distances. Then there exists $u\in V$ with $d(u) \neq d'(u)$. At least one of $d(u)$ and $d'(u)$ is in $\N$; therefore, we have $u \in W_i$ which shows that both $d(u)$ and $d'(u)$ are in $\N$. Among all $u\in V$ with $d(u) \neq d'(u)$, let $n$ be minimal such that either $d(u) > n = d'(u)$ or $d'(u) > n = d(u)$. Without loss of generality, suppose that $d'(u) > n = d(u)$. We have $u \not\in R$ because $d'(u) > 0$. If $u \in V_i \setminus R$, then there exists an edge $(u,v) \in E$ with $n-1 = d(v) = d'(v)$; the latter equality holds by minimality of $n$. This edge yields $d'(u) \leq n$, a contradiction. Let now $u \in V_{1-i}$, then all edges $(u,v) \in E$ satisfy $d(v) \leq n-1$ and hence $d(v) = d'(v)$, again by minimality of $n$. Note that there exists at least one such edge since $\max \emptyset = \infty$ but $d(u) < \infty$. As before, this shows $d'(u) \leq n$, a contradiction. Therefore $d(u) \neq d'(u)$ is not possible.
  \qed
\end{proof}

The following example shows that the second claim of Lemma~\ref{lem:windist} does not directly hold for infinite graphs. For finite game graphs, attractors are often defined as $\bigcup_{k \geq 0} \set{v \in V}{d(v) \leq k}$; see e.g.~\cite[p.145]{zielonka98tcs}. The example also shows that this approach does not work directly for graphs where positions can have infinitely many successors. Depending on the purpose, ordinal numbers might be used for a generalization of the winning distance towards infinite graphs.

\begin{example}
  Let $V = V_1 = \N \cup \os{a,b}$, i.e., all positions belong to player $1$. Let $R = \os{0} \subseteq \N$ be the target set of player $0$. The edges are $\set{(i+1,i) \in \N^2}{i \geq 0} \cup \set{(b,i)}{i \in \N} \cup \smallset{(a,b)}$. 
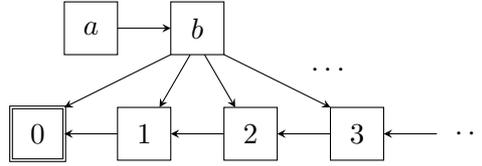
  \begin{center}
  \begin{tikzpicture}[scale=1.4]
    \node[rectangle,positionStyle] (p) at (0.5,1) {$a$};
    \node[rectangle,positionStyle] (q) at (1.5,1) {$b$};
    \node[rectangle,double,positionStyle] (0) at (0,0) {$0$};
    \node[rectangle,positionStyle] (1) at (1,0) {$1$};
    \node[rectangle,positionStyle] (2) at (2,0) {$2$};
    \node[rectangle,positionStyle] (3) at (3,0) {$3$};
    \node[rectangle,minimum size=\msize,inner sep=\isep,outer sep=\osep] (4) at (4,0) {};
    \node (5) at (4.1,0) {$\cdots$};
    \draw[-stealth] (p) -- (q);
    \draw[-stealth] (q.225) -- (0.45);
    \draw[-stealth] (q.255) -- (1);
    \draw[-stealth] (q.285) -- (2);
    \draw[-stealth] (q.315) -- (3.135);
    \draw[-stealth] (1) -- (0);
    \draw[-stealth] (2) -- (1);
    \draw[-stealth] (3) -- (2);
    \draw[-stealth] (4) -- (3);
    \draw (2.75,0.6) node {$\cdots$};
  \end{tikzpicture}
  \end{center}
  All positions in $V$ are winning for player $0$: all paths eventually end in $0$ because~$\N$ is well-ordered. The winning distance of the vertex $n \in \N$ is $d(n) = n$; in particular, the winning distances of the successors of $b$ are unbounded. Therefore, the winning distance of $b$ cannot be a natural number. Also note that the winning distance of~$a$ would need to be greater than $d(b)$.
  \eex
\end{example}

\begin{lemma}\label{lem:movesdist}
  Let $(W_i,F_i)$ be a maximal $i$-strategy for a reachability game in a finite game graph and let $d : V \to \N \cup \smallset{\infty}$ be the winning distance. For every $u \in W_i$, there exists a game $\alpha$ starting in $u$ and following $(W_i,F_i)$ which uses at least $d(u)$ moves.
\end{lemma}

\begin{proof}
  This is trivial if $d(u) = 0$. Let now $d(u) > 0$; in particular $u \not\in R$. First, consider the case $u \in V_i$. Let $(u,v) \in F_i$. Then $1 + d(v) \geq d(u)$ and, by induction, there exists a game $\beta$ starting at $v$ and following $(W_i,F_i)$ such that $\beta$ uses at least $d(v)$ moves. Then $\alpha = u\beta$ is a game which starts at $u$ and which follows $(W_i,F_i)$ and uses at least $1 + d(v) \geq d(u)$ moves.
  
  Next, let $u \in V_{1-i}$. Then there exists at least one edge $(u,v) \in E$ because $u \in W_i \setminus R$. Among the neighbors of $u$, we choose $v$ with $d(u) = 1 + d(v)$; the position $v$ exists by definition of $d(u)$. By induction, there exists a game $\beta$ starting at $v$ and following $(W_i,F_i)$ such that $\beta$ uses at least $d(v)$ moves; as before, $\alpha = u\beta$ is the desired game.
  \qed
\end{proof}

When trying to optimize the worst-case number of moves necessary to reach~$R$, then Lemma~\ref{lem:movesdist} shows that one cannot be better than the winning distance.
A maximal $i$-strategy $\sigma : W_i \to W_i$ actually achieves this bound if and only if for all $u \in W_i \cap V_i$ we have that $\sigma(u) = v$ implies $d(u) = 1 + d(v)$. In this case, we say that $\sigma$ is \emph{optimal}. For player $1-i$, every maximal winning strategy is optimal.

\begin{proposition}\label{prp:attralgo}
  Consider a reachability game with $n$ vertices and $m$ edges. Then we can compute optimal positional winning strategies for both players in time $\Oh(n + m)$. In particular, this computation yields attractors.
\end{proposition}

\begin{proof}
  We basically use an adaption of the breadth-first search algorithm (and if all positions belong to player $i$, then it actually is the usual breadth-first search algorithm; see e.g.~\cite[Chapter~20.2]{CLRS22}).
  Our algorithm uses the following data structures:
  \begin{itemize}
  \item A set $\cP$ which is initialized as $\cP = \set{ \big(\set{u\in V}{(u,v)\in E},v\big) }{ v \in V }$. For every $v \in V$, the set $\cP$ gives access to its predecessors. We assume that for a given $v \in V$, we have access to the pair $\big(U,v\big) \in \cP$ in constant time. We will successively remove edges in $\cP$ such that the remaining edges define strategies for the respective players. If we say that we remove an edge $(u,v)$ from $\cP$, then what we actually do is replacing the pair $(U,v) \in \cP$ by $(U\setminus{u},v)$.
  \item A function $n : V_{1-i} \to \N$ which gives the number of neighbors (i.e., successors, not predecessors) of a vertex in $V_{1-i}$ in the graph defined by $\cP$. Initially, we let $n(u)$ be the out-degree of $u$.
  \item A function $D : V \to \N \cup \smallset{\infty}$ which is the current estimate of the winning distance $d$. Initially, we have $D(u) = 0$ for $u \in R$ and $D(u) = \infty$ for $u \not\in R$. For each vertex $u$, the value $D(u)$ is assigned a new value at most once; if such an assignment occurs, then before this assignment, we have $D(u) = \infty$ and after this assignment we have $D(u) = d(u) < \infty$. If at some point we have $D(u) < \infty$ for $u \in V_{1-i} \setminus R$, then $n(u) = 0$.
  \item A FIFO queue $\cQ$ of vertices in $V$. Initially, $\cQ$ contains the vertices in $R$ in some arbitrary order. The queue $\cQ$ contains the vertices which still need to propagate their distance $D$ to their predecessors. An invariant of $\cQ$ will be that it only contains vertices $v$ with $D(v) = d(v) < \infty$ and that vertices with smaller winning distance are closer to the front of the queue than vertices with larger winning distance.
  \end{itemize}
  After this initialization, the algorithm proceeds as follows. While $\cQ\neq \emptyset$~do:
  \begin{enumerate}
  \item $v \gets \text{delete-first}(\cQ)$ and let $(U,v) \in \cP$.
  \item\label{stp:two} For all $u \in U$ do
    \begin{itemize}
    \item If $u \in V_i$, then
      \begin{enumerate}
      \item If $D(u) = \infty$, then $D(u) \gets 1+D(v)$ and append $u$ to $\cQ$;
      \item else we remove the edge $(u,v)$ from $\cP$.
      \end{enumerate}
    \item If $u \in V_{1-i}$, then 
      \begin{enumerate}[start=3]
      \item\label{stp:ccc} We remove the edge $(u,v)$ from $\cP$ and set $n(u) \gets n(u) - 1$.
      \item\label{stp:ddd} If $n(u) = 0$, then $D(u) \gets 1+D(v)$ and append $u$ to $\cQ$.
      \end{enumerate}
    \end{itemize}
  \end{enumerate}
  For $u \in V_i$, we set $D(u)= 1+D(v)$ when considering the first edge $(u,v)$; and for $u \in V_{1-i}$, we set $D(u)= 1+D(v)$ when considering the last remaining edge $(u,v)$. In both cases, the invariant on the order of the elements in $\cQ$ ensures that $D(u) = d(u)$.
  In step \ref{stp:ddd}, if $n(u) = 0$, we could remember the move $(u,v)$  since, even though it is losing for player $1-i$, always choosing these moves achieves a maximal winning distance for player $i$ (it might be a natural desire of player $1-i$ to delay the defeat for as long as possible).
  
  Every vertex $v \in V$ is added at most once to the queue $\cQ$. In the sum of all iterations of the loop in step \ref{stp:two}, we consider every edge of the graph at most once. Since the initialization is also possible in linear time, the running time of the above algorithm is $\Oh(n + m)$.
  
  After running the algorithm, $D=d$ is the winning distance and thus the winning positions are $W_i = \set{u \in V}{D(u) < \infty}$ and $W_{1-i} = \set{u \in V}{D(u) = \infty}$. The winning strategy $\sigma_i : W_i \to W_i$ for player $i$ at a position $u \in (W_i \cap V_i) \setminus R$ is given by $\sigma_i(u) = v$ with $(u,v) \in E$ and $D(u) = 1+D(v)$; note that in this case, after the algorithm stops, there exists exactly one pair $(U,v) \in \cP$ such that $D(v) < \infty$ and $u \in U$. The winning strategy $\sigma_{1-i} : W_{1-i} \to W_{1-i}$ for player $1-i$ at a non-sink position $u \in W_{1-i} \cap V_{1-i}$ is given by $\sigma_{1-i}(u) = v$ with $(u,v) \in E$ such that $v \in W_{1-i}$; note that if $u$ is not a sink, then there exists at least one such edge because $n(u) > 0$. Moreover, after the algorithm terminates, every such edge is represented by a pair $(U,v) \in \cP$ with $u \in U$. In other words, winning strategies for both player $i$ and $1-i$ are given by the edges in $\cP$; however, for player $i$, we need to exclude the edges leading to positions outside $W_i$.
  \qed
\end{proof}

\section{Parity Games}
The game graph $G=(V_0,V_1,E,\chi)$ of a \emph{parity game} is equipped with a vertex coloring $\chi : V \to \os{1,\ldots,d}$ for an integer $d \geq 1$. The coloring $\chi$ helps to formulate the winning conditions of the players. Sometimes, if we prefer the smallest color to be even, the coloring has the form $\chi : V \to \os{0,\ldots,d-1}$. The identifier $d$ is for \emph{dimension}. In the literature, the colors are called \emph{priorities}, too.
The subgraph of $G$ \emph{induced} by a set of vertices $W \subseteq V$ is $G[W] = (V_0 \cap W, V_1 \cap W, E', \chi')$ with $E' = \set{(u,v) \in E}{u,v \in W}$ and $\chi' : W \to \os{1,\ldots,d}$ is the restriction of $\chi$.
Similarly, $G-W$ is the subgraph of $G$ induced by $V \setminus W$.
In a parity game, player $0$ is called \Even and player $1$ is called \Odd. The set of games $C$ contains all infinite paths and all finite paths ending in a sink.
\Even wins all finite games which end in a sink in $V_1$ and infinite games where the largest color which is seen infinitely often is even. Symmetrically, \Odd wins all finite games which end in a sink in $V_0$ and all infinite games where the largest color which is seen infinitely often is odd. The winning condition regarding sinks means that if players cannot move, they lose immediately.

Whenever there are two colors $q$ and $q+2$ such that there is no position with color $q+1$, then we can identify $q$ and $q+2$ (for instance, by using the color $q$ for all vertices with color $q+2$). In particular, we can assume that the dimension~$d$ is the number of different colors in the game graph.

\begin{remark}\label{rem:edgecol}
In Section~\ref{sec:lehtinen}, we consider edge colorings for parity games. In this setting, a game graph has the form $G = (V_0,V_1,E,\chi)$ with $\chi:E\to \os{1,\ldots,d}$. As before, if the owner of a position cannot move, the owner loses immediately. Otherwise, in an infinite game player $i$ wins if the largest number $q$, which is seen infinitely often on the edges, satisfies $q \equiv i \bmod 2$. 
A parity game with vertex coloring $\chi$ can be transformed into a game with  edge coloring $\chi'$ simply by defining $\chi'(e)$ by $\chi(u)$ if $u$ is the source of $e$. For the other direction 
we introduce a smallest color $0$, and we subdivide every edge $e = (u,v)$ into a path $u \to v_e \to v$ for a new vertex $v_e$. The old vertices are colored with $0$ and the color of~$v_e$ is the color of $e$.
\eex
\end{remark}

\begin{remark}\label{rem:reachparity}
Reachability games can be encoded as parity games with two colors. Let $G=(V_0,V_1,E)$ be the game graph of a reachability game where (w.l.o.g.) it is player $0$'s objective to reach $R \subseteq V$. For all $v \in R$, we remove all outgoing edges. Then for all sinks $v$ (which now includes all positions in $R$), we introduce a self-loop $(v,v)$. After these two modifications, the resulting edge set is called~$E'$. We let $\chi(v) = 2$ if $v \in R$; otherwise, we set $\chi(v) = 1$. The parity game $G'=(V_0,V_1,E',\chi)$ now has the following property: player~$0$ wins the reachability game $(G,v)$ with target set $R$ if and only if player~$0$ wins the parity game $(G',v)$.
\eex
\end{remark}

As noticed in Remark~\ref{rem:reachparity}, we can eliminate all sinks in a parity game by introducing self-loops: 
let $G=(V_0,V_1,E,\chi)$ be a parity game with vertex coloring $\chi : V \to \os{1,\ldots,d}$. Let $S_i \subseteq V_i$ be the sinks belonging to player $i$. For every position $v \in S_0 \cup S_1$, we introduce a self-loop $(v,v)$. The resulting edge set is called~$E'$. Let $\chi' : V \to \os{1,\ldots, \max\os {2,d}}$ be a re-coloring of the vertices with 
  \begin{equation*}
    \chi'(v) = \begin{cases}
      i+1 & \ \text{ if } v \in S_{i} \text{ for } i \in \os{0,1}, \\
      \chi(v) & \ \text{ otherwise.}
    \end{cases}
  \end{equation*}
Then $(G,v)$ and $(G',v)$ for $G'=(V_0,V_1,E',\chi')$ have the same winner. The advantage is that the game graph $G'$ is without sinks. However, this construction introduces new edges and for $d=1$ it increases the number of colors, in general. 

A long sequence of results culminated in the following theorem~\cite{GurevichH82stoc,EmersonJ1991focs,Mostowski1991tr,mcnaughton1993,zielonka98tcs}.

\begin{theorem}\label{thm:paritydet}
  Parity games are positionally determined.
\end{theorem}

\begin{proof}
  Let $G = (V_0, V_1, E, \chi)$ be the game graph of a parity game with coloring $\chi : V \to \os{1,\ldots,d}$. We proceed by induction on $d$. If $d=1$, then \Even wins at all positions in $\attr_0(G,S_1)$ where $S_1$ are the sinks in $V_1$; all other positions are winning for \Odd. Let now $d>1$ and suppose that $d \equiv i \bmod 2$ for $i \in \os{0,1}$. Let $W_{1-i}$ be the support of a maximal $(1-i)$-strategy and let $W_i = V \setminus W_{1-i}$. We need to show that there exists an $i$-strategy with support $W_i$. The sinks in $V_i$ are all in $W_{1-i}$. On positions in $\attr_{1-i}(G,W_{1-i}) \setminus W_{1-i}$, player $1-i$ can force the game to visit a position in $W_{1-i}$ (see Proposition~\ref{prp:attr}) and from there on, player $1-i$ can follow the $(1-i)$-strategy with support $W_{1-i}$. Therefore, $\attr_{1-i}(G,W_{1-i})$ is the support of a winning strategy and, by maximality of $W_{1-i}$, we have $W_{1-i} = \attr_{1-i}(G,W_{1-i})$. It follows that all outgoing edges of positions in $W_i \cap V_{1-i}$ lead to positions in~$W_i$, and every position in $W_i \cap V_i$ has at least one outgoing edge to a position in~$W_i$.
  
  Let $H$ be the subgraph of $G$ induced by $W_i$, let $U_d = \set{v\in W_i}{\chi(v)=d}$, let $A = \attr_i(H,U_d)$ and let $G'$ be the subgraph of $H$ induced by $W_i \setminus A$.
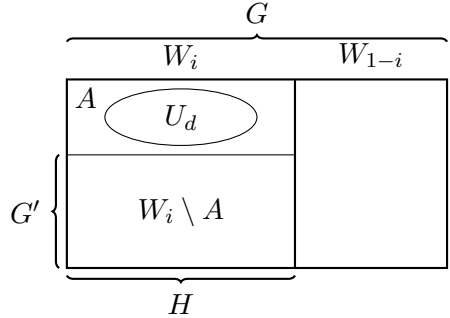
  \begin{center}
  \begin{tikzpicture}[xscale=0.5,yscale=0.5]
    \draw[thick] (0,0) rectangle (10,5);
    \draw[thick] (6,0) -- (6,5);
    \draw (0,3) -- (6,3);
    \draw (3,4) ellipse (2 and 0.75);
    \draw (0.5,4.5) node {$A$};
    \draw (3,4) node {$U_d$};
    \draw (3,1.5) node {$W_i \setminus A$};
    \draw (3,5) node[above] {$W_i$};
    \draw (8,5) node[above] {$W_{1-i}$};
    \draw [thick,decoration={brace},decorate]
      (6,-0.2) -- (0,-0.2) 
      node [pos=0.5,below,yshift=-0.1cm] {$H$};
    \draw [thick,decoration={brace},decorate]
      (-0.2,0) -- (-0.2,3)
      node [pos=0.5,left,xshift=-0.1cm] {$G'$};
    \draw [thick,decoration={brace},decorate]
      (0,6.0) -- (10,6.0) 
      node [pos=0.5,above,yshift=0.1cm] {$G$};
  \end{tikzpicture}
  \end{center}
   A $(1-i)$-strategy with nonempty support $W_{1-i}'$ on the game graph $G'$ yields a $(1-i)$-strategy with support $W_{1-i} \cup W_{1-i}'$ on $G$. Therefore, there is no nonempty $(1-i)$-strategy on $G'$. By induction on the number of colors, there exists a maximal $i$-strategy $\sigma'$ on $G'$ with support $W_i \setminus A$. This leads to the following strategy for player $i$ on positions in $W_i$ within the game graph $G$:
  \begin{itemize}
  \item At positions in $U_d \cap V_i$, player $i$ moves to a position in~$W_i$.
  \item At positions in $(A \setminus U_d) \cap V_i$, player $i$ moves according to the positional strategy for reaching $U_d$; see Proposition~\ref{prp:attr}.
  \item At positions in $(W_i \setminus A) \cap V_i$, player $i$ moves according to $\sigma'$.
  \end{itemize}
  The rules above define a positional strategy $\sigma$ for player $i$ with support $W_i$. To see that it is winning, consider any game $\alpha$ starting at a position in $W_i$ which follows $\sigma$. Since player $i$ never makes a move to $W_{1-i}$ and since player $1-i$ can never make a move to $W_{1-i}$, all positions of $\alpha$ are in $W_i$. If $\alpha$ is finite, then it ends in a sink in $V_{1-i}$ because there are no sinks in $V_i \cap W_i$; in particular, $\alpha$ is winning for player $i$ in this case. We can therefore assume that $\alpha$ is infinite. If $\alpha$ enters $A$ infinitely often, then $\alpha$ infinitely often visits a position in $U_d$. Therefore, the maximal color~$d$ is seen infinitely often; therefore, $\alpha$ is winning for player $i$. If, after some finite prefix, $\alpha$ stays in $G'$, then $\alpha$ is winning for player $i$ by choice of $\sigma'$. This shows that $\sigma$ is an $i$-strategy.
  \qed
\end{proof}

\begin{example}
  We consider the following parity game. Round vertices belong to \Even, square vertices belong to \Odd. The label $x{:}n$ means that the name of the vertex is $x$ and its color is~$n$.
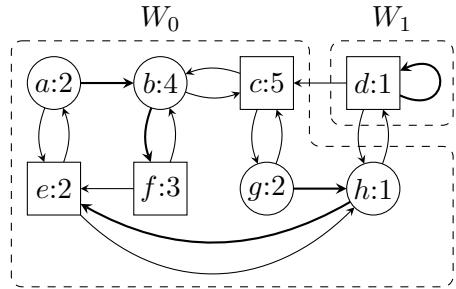
  \begin{center}
  \begin{tikzpicture}[scale=1.4]
    \node[circle,positionStyle] (a) at (0,1) {$a{:}2$};
    \node[circle,positionStyle] (b) at (1,1) {$b{:}4$};
    \node[rectangle,positionStyle] (c) at (2,1) {$c{:}5$};
    \node[rectangle,positionStyle] (d) at (3,1) {$d{:}1$};
    \node[rectangle,positionStyle] (e) at (0,0) {$e{:}2$};
    \node[rectangle,positionStyle] (f) at (1,0) {$f{:}3$};
    \node[circle,positionStyle] (g) at (2,0) {$g{:}2$};
    \node[circle,positionStyle] (h) at (3,0) {$h{:}1$};
    \draw[-stealth] (a) edge[bend right=20] (e);
    \draw[thick,-stealth] (a) edge (b);
    \draw[-stealth] (b) edge[bend right=20] (c);
    \draw[thick,-stealth] (b) edge[bend right=20] (f);
    \draw[-stealth] (c) edge[bend right=20] (b);
    \draw[-stealth] (c) edge[bend right=20] (g);
    \draw[-stealth] (d) edge (c);
	\draw[-stealth] (d) edge[bend right=20] (h);
    \draw[thick,-stealth] (d) edge[loop,out=-25,in=25,looseness=6] (d);
    \draw[-stealth] (e) edge[bend right=20] (a);
    \draw[-stealth] (e.315) .. controls (1,-1) and (2,-1) .. (h.225);
    \draw[-stealth] (f) edge (e);
    \draw[-stealth] (f) edge[bend right=20] (b);
    \draw[thick,-stealth] (g) edge (h);
    \draw[-stealth] (g) edge[bend right=20] (c);
    \draw[-stealth] (h) edge[bend right=20] (d);
    \draw[thick,-stealth] (h) edge[bend left=30] (e);
    \draw[dashed,rounded corners] (-0.4,1.4) -- (2.4,1.4) -- (2.4,0.4) -- (3.75,0.4) -- (3.75,-0.93) -- (-0.4,-0.93) -- cycle;
    \draw[dashed,rounded corners] (2.6,0.6) rectangle (3.75,1.4);
    \draw (3.175,1.4) node[above] {$W_1$};
    \draw (1,1.4) node[above] {$W_0$};
  \end{tikzpicture}
  \end{center}
  \Even's winning region is $W_0 = \os{a,b,c,e,f,g,h}$ and \Odd's winning region is $W_1 = \smallset{d}$. The respective winning strategies are indicated using thicker edges. We note that \Even's winning strategy is not unique since the strategy's move $(a,b)$ could be replaced by $(a,e)$.
  \eex
\end{example}

%
%

Let the maximal color satisfy $d \equiv i \bmod 2$ for $i \in \os{0,1}$.
The proof of Theorem~\ref{thm:paritydet} shows that as soon as we know the winning positions of player $1-i$, we can easily compute a winning strategy for player $i$.
However, the proof implicitly also gives an algorithm for computing the winning positions $W_{1-i}$ of player $1-i$. Let $X \subseteq V$ be minimal such that
\begin{itemize}
\item $X$ contains all sinks in $V_i$,
\item $X = \attr_{1-i}(G,X)$, and
\item (using the subgraph $G'$ from the proof of Theorem~\ref{thm:paritydet}) if $W_{1-i}'$ are the winning positions of player $1-i$ in the game $G'$, then $W_{1-i}'\subseteq X$.
\end{itemize}
The support $W_{1-i}$ of a maximal $(1-i)$-strategy satisfies the above properties; therefore we have $X \subseteq W_{1-i}$. On the other hand, the proof of Theorem~\ref{thm:paritydet} gives a winning strategy for player $i$ for all positions in $V \setminus X$. This yields $X = W_1$. In the next section, we consider this approach for solving finite parity games.

\section{Zielonka's Algorithm for Parity Games}\label{sec:Zielonka}

Let the number of positions $V$ in $G$ be finite and suppose that the largest color $d$ satisfies $d \equiv i \bmod 2$ for $i \in \os{0,1}$. We can assume that there is at least one vertex with color $d$; otherwise, we decrease the dimension $d$. We initialize $W_{1-i} = \attr_{1-i}(G,S_i)$ where $S_i$ are the sinks belonging to player $i$. Then we iterate the following steps until $W_{1-i}$ does not increase anymore (i.e., until $W_{1-i}' = \emptyset$ or $W_{1-i} = V$):
  \begin{enumerate}
  \item $W_i \gets V \setminus W_{1-i}$ and $H \gets G[W_i]$
  \item $U_d \gets \set{v\in W_i}{\chi(v)=d}$
  \item $A \gets \attr_i(H,U_d)$
  \item $G' \gets G[W_i \setminus A]$
  \item\label{stp:rec} Let $W_{1-i}'$ be the winning positions of \Odd in $G'$.
  \item\label{stp:union} $W_{1-i} \gets W_{1-i} \cup W_{1-i}'$
  \item\label{stp:attr} $W_{1-i} \gets \attr_{1-i}(G,W_{1-i})$
  \end{enumerate}
  Note that Step~\ref{stp:rec} consists of a recursive call for a game graph with fewer vertices (since there is at least one vertex $v \in V$ with $\chi(v) = d$) and colors $\os{1,\ldots,d-1}$. There is no recursive call if $G'$ is empty.

  After the algorithm terminates, the winning regions of the two players are $W_i$ and $W_{1-i}$; note that the correctness of the algorithm was proven in the previous section.
The above algorithm can also compute a $(1-i)$-strategy with support $W_{1-i}$: Initially, the strategy with support $W_{1-i}$ is the strategy for winning a reachability game. If $W_{1-i}$ is increased in Step~\ref{stp:union}, then we unite the two winning strategies (the strategy for $W_{1-i}$ before the assignment and the strategy for $W_{1-i}'$). If $W_{1-i}$ is increased in Step~\ref{stp:attr}, then on the new positions we play according to the strategy for reaching the positions in $W_{1-i}$ before this assignment.

The $i$-strategy with support $W_i$ can be computed as follows: During the last iteration of the loop, we have $W_{1-i}' = \emptyset$; moreover, we can assume that the recursive call in Step~\ref{stp:rec} also returns an $i$-strategy $\sigma'$ with support $W_i \setminus A$. As in the proof of Theorem~\ref{thm:paritydet}, a winning strategy for player $i$ is as follows: at position in $U_d$, player $i$ makes some arbitrary move to a position in $W_i$; at positions in $A \setminus U_d$, player $i$ moves according to the positional strategy for reaching $U_d$; and at positions in $W_0 \setminus A$, player $i$ moves according to $\sigma'$.
Without any significant additional effort, we can therefore assume that the above algorithm also computes the corresponding winning strategies.
This approach for computing maximal positional winning strategies is known as \emph{Zielonka's algorithm}.

\begin{theorem}
  Let $G$ be a parity game with $n$ vertices, $m$ edges, and $d$ colors. Then Zielonka's algorithm computes maximal winning strategies for both players in time $\Oh\big(n^{d-1}(n + m)\big)$ and, thus, in time $\Oh\big(n^{d+1}\big)$.
\end{theorem}

\begin{proof}
  Let $c \geq 1$ be a constant such that the initialization and one iteration of the loop when omitting the time for the recursive call in Step~\ref{stp:rec} takes time at most $c \cdot (n+m)$; see Proposition~\ref{prp:attralgo}. Let $f(n,d) \cdot c \cdot (n+m)$ be the running time of the algorithm. It suffices to show that $f(n,1) \leq 1$ and $f(n,d) \leq 2n^{d-1}$ for $d \geq 2$.  This is true for $d=1$ because there is no recursive call. Let now $d \geq 2$.
  The game graph $G'$ is always smaller than $G$. Since the size of $H$ decreases with every iteration, there are at most $n$ iterations of the loop. Every recursive call in Step~\ref{stp:rec} uses at most $n-1$ positions and $d-1$ colors. Hence, the worst-case running time satisfies
  \begin{equation*}
    f(n,d) \cdot c \cdot (n+m) \leq n \cdot \big(f(n-1,d-1) \cdot c \cdot (n+m) + c \cdot (n+m)\big)
  \end{equation*}
  Dividing by $c \cdot (n+m)$ yields
  \begin{equation}
    f(n,d) \leq n \cdot \big(f(n-1,d-1) + 1\big)
    \label{equ:zie}
  \end{equation}
  For $d=2$, we obtain $f(n,2) \leq n \big(f(n-1,1) + 1\big) \leq 2n = 2n^{d-1}$ because $f(n-1,1) \leq 1$. Let now $d>2$. Using Equation~\eqref{equ:zie}, we see that
  \begin{align*}
    f(n,d) 
    &\leq n \cdot \big( 2(n-1)^{d-2} + 1 \big) \tag*{\text{by induction hypothesis}}
    \\ &\leq n \cdot \big( 2(n-1)n^{d-3} + 1 \big) \tag*{\text{since $d\geq 3$}}
    \\ &= n \cdot \big(2n^{d-2}-2n^{d-3} + 1 \big)
    \\ &\leq 2n^{d-1} \tag*{\text{since $-2n^{d-3}+1 < 0$}}
  \end{align*}
  The second part of the statement follows since $\Oh(n+m) \subseteq \Oh(n^2)$.
  \qed
\end{proof}

\begin{example}
  Consider the following parity game with $2n$ vertices and $2$ colors.
  \Even's positions are $V_0 = \os{a_1,\ldots,a_n}$ and \Odd's positions are $V_1 = \os{b_1,\ldots,b_n}$. All vertices in $V_0$ have the color $1$ and all vertices in $V_1$ have the color $2$. We have loops $(a_i,a_i)$ and edges $(b_i, a_i)$ for all $i$ as well as edges $(a_i,b_j)$ for all $i < j$. All positions are winning for \Odd. The game graph for $n=4$ is:
  
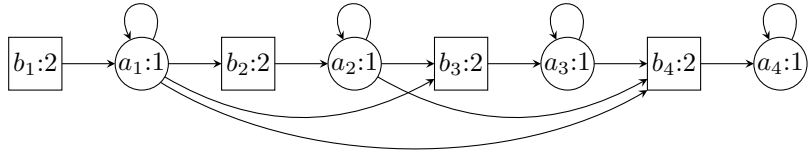
\begin{center}
  \begin{tikzpicture}[scale=1.4]
    \node[rectangle,positionStyle] (b1) at (0,0) {\small$b_1{:}2$};
    \node[circle,positionStyle] (a1) at (1,0) {\small$a_1{:}1$};
    \draw[-stealth] (b1) edge (a1);
    \draw[-stealth] (a1) edge[loop,out=65,in=115,looseness=6] (a1);
    
    \node[rectangle,positionStyle] (b2) at (2,0) {\small$b_2{:}2$};
    \node[circle,positionStyle] (a2) at (3,0) {\small$a_2{:}1$};
    \draw[-stealth] (b2) edge (a2);
    \draw[-stealth] (a2) edge[loop,out=65,in=115,looseness=6] (a2);
    
    \node[rectangle,positionStyle] (b3) at (4,0) {\small$b_3{:}2$};
    \node[circle,positionStyle] (a3) at (5,0) {\small$a_3{:}1$};
    \draw[-stealth] (b3) edge (a3);
    \draw[-stealth] (a3) edge[loop,out=65,in=115,looseness=6] (a3);
    
    \node[rectangle,positionStyle] (b4) at (6,0) {\small$b_4{:}2$};
    \node[circle,positionStyle] (a4) at (7,0) {\small$a_4{:}1$};
    \draw[-stealth] (b4) edge (a4);
    \draw[-stealth] (a4) edge[loop,out=65,in=115,looseness=6] (a4);

    \draw[-stealth] (a1) edge (b2);
    \draw[-stealth] (a1) edge[bend right] (b3);
    \draw[-stealth] (a1.315) .. controls (2.5,-1) and (4.5,-1) .. (b4.225);
    
    \draw[-stealth] (a2) edge (b3);
    \draw[-stealth] (a2) edge[bend right] (b4);
    
    \draw[-stealth] (a3) edge (b4);
  \end{tikzpicture}
  \end{center}
  Initially, we have $W_1 = \emptyset$.
  In the first iteration of Zielonka's algorithm, we compute the $0$-attractor of $U_2 = V_1$ which is $V \setminus \smallset{a_n}$. This computation uses a quadratic number of steps; see Proposition~\ref{prp:attralgo}. Then the recursive call returns $W_1' = \smallset{a_n}$, after which we have $W_1 = \os{a_n,b_n}$ by computing the $1$-attractor. The next iteration is similar, but with $n$ decreased by $1$. Since we have $n$ iterations, this yields a cubic running time of Zielonka's algorithm. For $d=2$, this shows that the bound of $\Oh(n^3)$ on the running time of Zielonka's algorithm is tight.
  \eex
\end{example}

\section{Lehtinen's Algorithm for Parity Games}\label{sec:lehtinen}

The idea of Lehtinen's algorithm is to translate a given parity game into another parity game such that solutions of the new game yield solutions of the original game. Moreover, applying Zielonka's algorithm to the new parity game yields a quasi-polynomial running time $2^{\log^{\Oh(1)}n}$.


We need the following notions for graphs. A nonempty set of vertices $U \subseteq V$ is \emph{strongly connected} if for all $u,v \in U$ there exists a path from $u$ to $v$. Every singleton subset $\smallset{v} \subseteq V$ is strongly connected. A \emph{strongly connected component} is a maximal strongly connected subset; i.e., $U$ is a strongly connected component if there is no strongly connected subset $U'$ with $U \subsetneq U' \subseteq V$. Every graph can be partitioned into strongly connected components; if $U$ and $U'$ are different strongly connected components, then there cannot exist paths both from $U$ to $U'$ and from $U'$ to $U$. Therefore, by successively moving from one strongly connected component to another strongly connected component, one cannot visit the same strongly connected component twice. In particular, in a finite graph, there exist strongly connected components $U$ such that one cannot reach any other strongly connected; in this case $U$ is called \emph{terminal}.

Let $G = (V_0,V_1,E,\chi)$ be the game graph of a parity game with vertices $V = V_0 \cup V_1$ and vertex coloring $\chi : V \to \os{1,\ldots,d}$. 
The $r$-register graph $R^r(G)$ of $G$ is again a game graph, but with an edge coloring, see Remark~\ref{rem:edgecol}.
To avoid confusion, the vertices of $R^r(G)$ are called states.
The states are the elements $(v,x,p) \in V \times \N^{r} \times \{s,t\}$ where $x = x_1 \cdots x_r$ satisfies $x_1 \leq \cdots \leq x_r$. States with $p=s$ are called \emph{reset states}; states with $p=t$ are called \emph{transition states}.
All reset states belong to \Even, every transition state $(v,x,t)$ belongs to the owner of $v$.
For every register $j \in \{1,\ldots,r\}$ and every reset state $(v,x,s)$, there is an outgoing edge with label $reset(j)$. The target state is $(v,y,t)$ with $y = (0,x_1,\ldots,x_{j-1},x_{j+1},\ldots,x_r)$. Formally, the label of the edge is not important it helps with reasoning about the game.
If there is an outgoing edge at a transition state $(v,x,t)$, then its target is a reset state $(w,y,s)$ with $(v,w)\in E$ and $y_j = \max(\chi(w),x_j)$.
The first kind of edges are called \emph{resets} and the second kind of edges are called \emph{transitions}.
All paths alternate between resets and transitions.
The color of all transitions is $0$; the color of an edge with label $\reset(j)$ is $2j$ if the value $x_j$ of register $j$ before the reset is even and $2j+1$ if $x_j$ is odd.
The $r$-register game $R^r(G,v,x)$ is the parity game with initial state $(v,x,s)$; its game graph is the subgraph of $R^r(G)$ induced by the states reachable from $(v,x,s)$.
For every game $\alpha$ in $R^r(G)$ starting at a state $(u,x,p)$ there exists a \emph{corresponding} game $\alpha_G$ in $G$ starting at $u$. The game $\alpha_G$ is the sequence of first components at the transition states of $\alpha$.
Note that register games are not symmetric for the two players: Firstly, all reset states belong to \Even. And secondly, resets of register $j$ can have even and odd colors, but the odd color $2j+1 $ is larger than the corresponding even color $2j$.
It is this second property which leads to the following lemma.

\begin{lemma}\label{lem:odd2odd}
  Let $r \geq 1$, let $\alpha$ be a game in $R^r(G)$ and let $\alpha_G$ be the corresponding game in $G$. If 
  $\alpha_G$ is winning for \Odd, then so is $\alpha$.
\end{lemma}

\begin{proof}
  Let $\alpha_G$ be winning for \Odd.
  The game $\alpha$ ends in the sink $(v,x,t)$ if and only if $\alpha_G$ ends in the sink $v$. In particular, both sinks $(v,x,t)$ and $v$ then belong to the same player. Also note that reset states cannot be sinks. Since $\alpha_G$ is winning for \Odd, we can assume that $\alpha$ is infinite (otherwise, $\alpha$ would end in a sink belonging to \Even and therefore be winning for \Odd, as desired). 

  Let $q$ be the largest color which is seen infinitely often during the game $\alpha_G$, and let $j$ be the largest register such that \Even infinitely often plays $\reset(j)$ in the game $\alpha$. There is a point in $\alpha_G$ after which no color larger than $q$ occurs. At the corresponding point in $\alpha$, we can wait for $j$ resets of register $j$. From then onwards, the value $x_j$ of register $j$ is at most $q$. In particular, whenever we then see the color $q$ in $\alpha_G$, the contents of register $j$ is $q$, and it stays $q$ at least until the next reset of register $j$. Therefore, there are infinitely many resets of register $j$ when its value is $q$. Since $\alpha_G$ is winning for \Odd, the number $q$ is odd. We therefore infinitely often see the number $2j+1$ in the colors of $\alpha$. Since all larger registers are reset only finitely often, $2j+1$ is the largest color of $\alpha$ which is seen infinitely often. Therefore, $\alpha$ is winning for \Odd.
  \qed
\end{proof}

\begin{lemma}\label{lem:oddwins}
  If \Odd wins $(G,v)$, then \Odd wins $R^r(G,v,x)$ for all $r \geq 1$ and all $x \in \N^r$.
\end{lemma}

\begin{proof}
  If \Odd wins $(G,v)$, then there exists a positional strategy $(V,F)$ such that \Odd wins $(G,v)$ by following this strategy. We adapt this strategy to $R^r(G)$: at a state $(u,y,t)$ with $u \in V_1$, \Odd moves to $(u',y',s)$ with $(u,u') \in F$. It remains to show that this strategy is winning for \Odd in the register game $R^r(G,v,x)$. Let $\alpha$ be a game starting at $(v,x,s)$ and following the above strategy. The corresponding game $\alpha_G$ in $G$ starts at $v$ and follows the strategy $(V,F)$. Therefore $\alpha_G$ is winning for \Odd. By Lemma~\ref{lem:odd2odd}, \Odd wins the game $\alpha$.
  \qed 
\end{proof}

\begin{remark}\label{rem:oddwinsContrapos}
  Positional determinacy of parity games (Theorem~\ref{thm:paritydet}) leads to the following consequence of Lemma~\ref{lem:oddwins}: if \Even wins  $R^r(G,v,x)$ for some $r \geq 1$ and $x \in \N^r$, then \Even wins $(G,v)$. When following an analogous approach as above, then a direct proof for this consequence would need to translate a winning strategy for the register game $R^r(G,v,x)$ into a winning strategy for $(G,v)$. However, even if \Even's winning strategy for $R^r(G,v,x)$ is positional, the resulting strategy for $(G,v)$ might not be positional because some different contents of the registers could lead to different moves at a given position in $G$.
  \eex
\end{remark}

For a weak converse of Lemma~\ref{lem:oddwins}, we will use induction on the number of vertices. During this induction, smaller register games occur. Here, ``smaller'' either refers to the size of the corresponding game graph $G$ or the number of registers.
If $G'$ is a subgraph of $G$, then $R^r(G')$ is a subgraph of $R^r(G)$.
If $q\leq r$, then every positional strategy on $R^{q}(G)$ for \Even defines a strategy on $R^r(G)$ in which \Even never plays $\reset(j)$ for $j>q$.

During the proof of the following proposition, we will use a slightly different notion of a positional strategy. Instead of just the support, we assume that a positional strategy $\sigma : V \to V$ for player $i$ is defined for all $v \in V_{i}$ which are not sinks (i.e., $\sigma$ is also defined for the positions outside the support). Moreover, the proof will use defensive strategies. A strategy for \Even in $R^r(G)$ is \emph{defensive} if $\reset(r)$ is never played when the contents of register $r$ is odd.

\begin{proposition}\label{prp:evenwins}
  If \Even wins $(G,v)$ and $\abs{V} < 2^r$, then \Even wins $R^r(G,v,x)$ for all $x \in \N^r$.
\end{proposition}

\begin{proof}
  We assume without restriction that all vertices in $G$ are reachable from $v$.
  We proceed by induction on $\abs{V}$. First, suppose that $V = \smallset{v}$. If $v$ is a sink, then it belongs to \Odd because \Even wins; in this case, after playing $\reset(1)$ at state $(v,x,s)$, the game in $R^r(G)$ also ends in a sink belonging to \Odd. If $v$ is not a sink, then there is a self-loop and the color $\chi(v)$ is even (again because \Even is wins). \Even again always plays $\reset(1)$ and after the first reset (in which case the color depends on $x$), the color of all resets is $2$. Since all other colors are $0$, \Even wins the register game. If $r=1$, then $\abs{V}=1$ and this case was already considered. In the remainder of the proof, we can therefore assume that $r>1$ and $\abs{V}>1$. 

We fix a positional winning strategy for \Even for $(G,v)$. At transition states in the register game, \Even always moves according to her strategy for $G$. We can remove all edges starting at \Even's positions in $G$ which are not part of the winning strategy. Since now, all reachable positions belonging to \Even have exactly one outgoing edge, we can transfer ownership of these positions to \Odd. In particular, now all transition states of $R^r(G,v,x)$ belong to \Odd and the game alternates between \Even's resets and \Odd's transitions.

We show that \Even can win $R^r(G,v,x)$ against every positional strategy of \Odd. Let $T$ be the transitions defining \Odd's strategy. 
Consider is a strongly connected subgraph $G'$ of $G$ and let $w$ be a vertex of $G'$ such that $(w,y,s)$ is a state of $R^r(G,v,x)$.
Then $R^r(G',w,y)$ is a subgraph of $R^r(G,v,x)$.
We say that a positional strategy of \Even for $R^r(G',w,y)$ \emph{is leaving} this subgraph if, starting at $(w,y,s)$ in $R^r(G,v,x)$, alternating between \Even's strategy for resets and \Odd's strategy $T$ for transitions eventually leads to a state outside $R^r(G',w,y)$. 

As before for the graph $G$ and \Even's strategy, we can remove all transitions from the register game except for those in \Odd's strategy $T$. We cannot transfer ownership because there might be sinks owned by \Odd. However, there is never any choice to be made by \Odd. \Even, starting from $(v,x,s)$, moves to some terminal strongly connected component $H$ of $R^r(G,v,x)$; this means that there is no other strongly connected component which is reachable from~$H$. In general, \Even's strategy for reaching $H$ is not defensive. The underlying positions (i.e., first components) of states in $H$ form a strongly connected subgraph~$G'$ of~$G$. If $G'$ consists of a single sink, then this sink belongs to \Odd (and, thus, \Even also wins the corresponding register game). We can therefore assume that $G'$ is not a sink. Since every position in $G'$ lies on some non-trivial loop, the largest color $d$ of $G'$ is even.
Since $H$ is a terminal strongly connected component, all registers have only values which appear in $G'$ (otherwise, \Even could move to another strongly connected component by playing $\reset(r)$). Next, \Even goes to one of the vertices $(w,y,s)$ such that $\chi(w)=d$. We now have $y=(d,\ldots,d)$; therefore, we can apply the following claim which then completes the proof of the proposition because $H$ cannot be left.

\medskip

\noindent
\emph{Claim: }
Let $G'$ be a strongly connected subgraph of $G$ which is not a single sink.
Let $d$ be the largest color of the vertices of $G'$. 
Let $w$ be a vertex of $G'$. 
Let $y=(y_1,\ldots,y_r)$ be such that $y_r \geq d$ and $y_r$ is even, and $y_j \leq d$ for all $j<r$. 
  Then \Even has a defensive strategy for $R^r(G',w,y)$ which is either winning or which is leaving $R^r(G',w,y)$.
  
\smallskip

\noindent
\emph{Proof of the claim: }
The color $d$ is even because \Even wins $(G,v)$.
The proof is by induction on the number of vertices in $G'$. 
If $G'$ contains a single vertex, then \Even always plays $\reset(1)$ with color $2$. This either yields an infinite game with the maximal infinite color being $2$, or it ends in a sink belonging to \Odd, or it leaves $R^r(G',w,y)$. In either case, the claim is true. Let now $G'$ have at least two vertices.
  Let $U_d$ consist of all vertices of $G'$ with color $d$.
  Let $G_1,\ldots,G_k$ be the strongly connected components of $G'-U_d$. Some of the components might be sinks -- even if $G'$ does not have any sinks. If $G_j$ has less than $2^{r-1}$ vertices, then the induction hypothesis in the proof of the proposition yields a winning strategy for the $(r-1)$-register game on $G_j$. 
  Since $G$ has less than $2^r$ vertices, at most one component has $\geq 2^{r-1}$ vertices. Let this component be $G_1$; (it might be that there is no such component, in which case we can simply assume that $G_1$ is never visited). It remains to give winning strategies for \Even for register games on $G_1$ and on $U_d$. Suppose that we have such strategies, then \Even always plays these winning strategies until \Odd's strategy~$T$ forces her to move to
  \begin{itemize}
  \item a register game on another strongly component $G_1,\ldots,G_k$,
  \item a register game on $U_d$, or
  \item to leave $R^r(G',v,x)$.
  \end{itemize}
  In the first two cases, \Even again applies the winning strategy of the corresponding register game. If the last case occurs, then the claim is true.
  Note that every path in $G'$ from $G_j$ to $G_\ell$ and back to $G_j$ with $j\neq \ell$ visits $U_d$ at some point before re-entering $G_j$.
  
  Next, we describe the winning strategies for the register games on $G_1$ and on $U_d$. We will maintain the following invariants: Firstly, whenever we are leaving $G_1$, the contents of register $r$ will either be even or less than $d$. And secondly, whenever we are leaving $U_d$, the contents of register $r$ will be $d$. All other registers always have values $\leq d$.
  The components $G_2,\ldots,G_k$ do not affect these invariants because the colors are $\leq d$ and the strategies never play $\reset(r)$.
  
  At reset states $(u,y,s)$ with $u \in U_d$, \Even plays $\reset(r)$ and moves to a transition state $(u,y',t)$. When entering $(u,y,s)$, the invariants ensure that the contents $y_r$ of register $r$ is an even number $\geq d$ and $y_j = d$ for all $j < r$. Therefore, the color of this reset edge is $2r$ and the value $y'_r$ of register $r$ after the reset is $y_{r-1} = d$ because $r \geq 2$.
    
  It remains to consider the component $G_1$. \Even plays $\reset(r-1)$ until all registers except for register $r$ have a value which occurs in $G_1$ or until we are leaving $G_1$, whichever happens first. If we are leaving $G_1$, then all colors of the resets are $\leq 2r-1$. Otherwise, the induction hypothesis in the proof of the claim yields a defensive winning strategy for the register game on $G_1$, and \Even continues by following this strategy. Note that all resets in following a defensive strategy have colors $\leq 2r$. \Odd's strategy $T$ might force us to leave $G_1$ with a register configuration where the top register is smaller than $d$ (and this value could even be over-written by another number smaller than $d$ in one of the other components $G_2,\ldots,G_k$) but it will again be $d$ when re-entering $G_1$ since we need to visit $U_d$ before re-entering.

A game which follows \Even's strategy and which does not leave $G'$ can fall into one of two categories. The first category is that it infinitely often visits states $(u,y,s)$ with $u \in U_d$. Then we have infinitely many resets with color $2r$ and all other resets have colors $\leq 2r$. Therefore, this game is winning for \Even. The second category is that after some time, the game stays in one of the components $G_j$ (i.e., all states $(u,y,p)$ after some finite prefix have first components $u$ in $G_j$). In this case, the game is winning because (after some finite prefix) it follows a strategy for $G_j$ which is winning for \Even.
This completes the proof of the claim and, hence, the proof of the proposition.
\qed
\end{proof}

\begin{theorem}[Lehtinen~\cite{Lehtinen18}]\label{thm:lehtinen}
  Let $G$ be a parity game with $n$ vertices and let $r \geq 1$ such that $n < 2^r$. Then for every position $v$ and every $x \in \N^r$, the games $(G,v)$ and $R^r(G,v,x)$ have the same winner.
\end{theorem}

\begin{proof}
  If \Odd wins $(G,v)$, then \Odd wins $R^r(G,v,x)$ by Lemma~\ref{lem:oddwins}. If \Even wins $(G,v)$, then \Even wins $R^r(G,v,x)$ by Proposition~\ref{prp:evenwins}. The claim follows by Theorem~\ref{thm:paritydet}.
  \qed
\end{proof}

\begin{remark}\label{rem:lehtlower}
Lehtinen shows that there are parity games with $n$ vertices such that \Even wins at every vertex but she needs $\Omega(\log n)$ registers to win the corresponding register game~\cite[Lemma~4.4]{Lehtinen18}. We revisit Lehtinen's example in terms of vertex colorings: We inductively construct a game graph $G_r$ with $2^{r+1}-2$ vertices all belonging to \Odd, with largest color $2r$, and such that there is exactly one position with color $2r$ and one with color $2r-1$. \Even wins $(G_r,v)$ at all starting positions $v$  but no register game $R^r(G_r,v,x)$. Moreover, \Even wins $R^{r+1}(G_r,v,x)$ for all $v \in V$ and all $x \in \N^{r+1}$. We let $R_1$ be the following game graph with colors $\os{1,2}$:
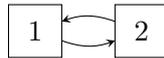
\begin{center}
\begin{tikzpicture}[scale=1.4]
  \node[rectangle,positionStyle] (a) at (0,0) {$1$};
  \node[rectangle,positionStyle] (b) at (1,0) {$2$};
  \draw[-stealth] (a) edge[bend right=20] (b);
  \draw[-stealth] (b) edge[bend right=20] (a);
\end{tikzpicture}
\end{center}
The graph $G_r$ for $r > 1$ is constructed from two copies of $G_{r-1}$ and two new vertices with colors $2r-1$ and $2r$, respectively:
  
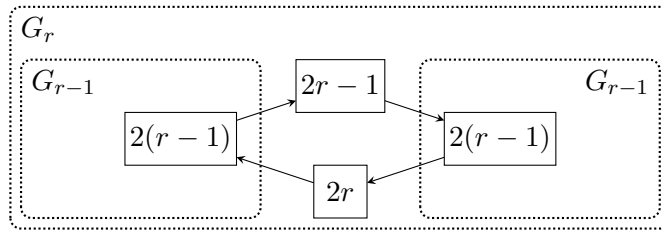
\begin{center}
  \begin{tikzpicture}[scale=1.4]
    \node[rectangle,positionStyle] (b) at (0.5,0) {$\,2(r-1)\,$};
    \node[rectangle,positionStyle] (c) at (2,0.5) {$\,2r-1\,$};
    \node[rectangle,positionStyle] (d) at (2,-0.5) {$\,2r\,$};
    \node[rectangle,positionStyle] (e) at (3.5,0) {$\,2(r-1)\,$};

    \draw[-stealth] (b) -- (c);
    \draw[-stealth] (c) -- (e);
    \draw[-stealth] (d) -- (b);
    \draw[-stealth] (e) -- (d);
    
    \draw[thick,densely dotted,rounded corners] (-1,-0.75) rectangle (1.25,0.75);
    \draw[thick,densely dotted,rounded corners] (2.75,-0.75) rectangle (5,0.75);
    \draw (-1,0.75) node[below right] {$G_{r-1}$};
    \draw (5,0.75) node[below left] {$G_{r-1}$};
    
    \draw[thick,densely dotted,rounded corners] (-1.1,-0.85) rectangle (5.1,1.25);
    \draw (-1.1,1.25) node[below right] {$G_r$};
  \end{tikzpicture}
  \end{center}
 We briefly describe \Odd's winning strategy for the $r$-register game in terms of the underlying graph. As long as \Even only resets registers~$\leq r-1$, \Odd always stays within one of graphs $G_{r-1}$. After \Even resets register~$r$ at least $r$ times, \Odd changes to the other copy of $G_{r-1}$ via one of the vertices with color $2r-1$ or $2r$. \Even wins the $(r+1)$-register game, basically by resetting register $j+1$ at positions with color $2j$ and register $1$ at all other positions.
  \eex
\end{remark}

\begin{theorem}[Calude, Jain, Khoussainov, Li, and Stephan \cite{CaludeJKLWS2017stoc}]\label{thm:CJKLS}
If $G$ is a parity game with $n$ vertices, then we can decide the winner of $(G,v)$ in quasi-polynomial time $2^{\log^{\Oh(1)}(n)}$.
\end{theorem}

\begin{proof}
  Let the vertex coloring be $\chi : V \to \os{1,\ldots,d}$. 
  Let $r \geq 1$ be minimal such that $n < 2^r$. Then $r \in \Oh(\log n)$. By Theorem~\ref{thm:lehtinen}, it suffices to solve $R^r(G,v,\mathbf{0})$ for $\mathbf{0} = (0,\ldots,0)$ in quasi-polynomial time to decide the winner of $(G,v)$. The positions of $R^r(G,v,\mathbf{0})$ are all in $V \times \os{0,\ldots,d}^r \times \os{s,t}$. In particular, there are at most $2n (d+1)^r$ positions in the register game with colors $\smallset{0} \cup \os{2,\ldots,2r+1}$. We can solve this game using Zielonka's algorithm in time $n^{\Oh(r)} d^{\Oh(r^2)}$, see Remark~\ref{rem:edgecol}. This yields a running time of $n^{\Oh(\log n)} d^{\Oh(\log^2 n)}$ since $r \in \Oh(\log n)$. We can assume that $d \leq n+1$ which yields a quasi-polynomial running time $n^{\Oh(\log^2 n)} = 2^{\Oh(\log^3(n))}$. 
  \qed
\end{proof}

\section{Conclusion}

In this survey paper, we revisit Lehtinen's quasi-polynomial algorithm for solving parity games~\cite{Lehtinen18}, and we provide all necessary preliminary results with full proofs. This includes the following topics:
\begin{itemize}
\item attractors and positional determinacy of reachability games,
\item the computation of optimal winning strategies for reachability games,
\item positional determinacy of parity games,
\item an analysis of Zielonka's algorithm for solving parity games,
\item and Lehtinen's register games.
\end{itemize}
Both determinacy results are proven for arbitrary game graphs; in particular, the graphs are allowed to be infinite. While reachability games can end after finitely many moves if the target set is reached, typical parity games have an infinite duration (except if they end in a sink). For a uniform treatment, we use a framework which includes both finite and infinite durations. 

It would be interesting to have a tighter analysis of Zielonka's algorithm. Friedmann gives a game with a linear number of vertices and colors such that the running time of Zielonka's algorithm on this game takes time at least $F_n$ for the $n$-th Fibonacci number~\cite{Friedmann2011rairo}. There is still a significant gap between this lower bound and the upper bound in Section~\ref{sec:Zielonka}. Whether finite parity games can be solved in polynomial time is still the main open problem is this area.

{\small
\newcommand{\Ju}{Ju}\newcommand{\Ph}{Ph}\newcommand{\Th}{Th}\newcommand{\Ch}{Ch}\newcommand{\Yu}{Yu}\newcommand{\Zh}{Zh}\newcommand{\St}{St}\newcommand{\curlybraces}[1]{\{#1\}}

}

\end{document}